\documentclass[aps,prl,floatfix,showpacs,twocolumn,pdffig]{revtex4}
\usepackage{amsmath,amssymb}
\usepackage{graphicx}
\usepackage{epsfig}
\usepackage{psfrag}
\usepackage{hyperref}

\newcommand{\be}{\begin{equation}}
\newcommand{\ee}{\end{equation}}

\begin{document}

\hsize\textwidth\columnwidth\hsize\csname@twocolumnfalse\endcsname

\title{Broken SU(4) Symmetry and The Fractional Quantum Hall Effect in Graphene} 

\author{I. Sodemann}
\author{A.~H. MacDonald}
\affiliation{Department of Physics, University of Texas at Austin, Austin TX 78712, USA}

\begin{abstract}
We describe a variational theory for incompressible ground states and charge gaps 
in the $N=0$ LL of graphene that accounts for the 4-fold Landau level degeneracy and the 
short-range interactions that break SU(4) spin-valley invariance. 
Our approach explains the experimental finding that gaps at odd numerators are weak 
for $1 < |\nu| < 2$ and strong for $0 < |\nu| < 1$.  We find that in the SU(4) invariant case
the incompressible ground state at $|\nu|=1/3$ is a three-component incompressible state, not the Laughlin state, 
and discuss the competition between these two states in the presence of SU(4) spin-valley symmetry breaking terms.   
\end{abstract}

% Electronic structure graphene, 73.22.Pr
% Fractional quantum Hall effect, 73.43.-f

\maketitle

\noindent
{\em Introduction}---The fractional quantum Hall effect (FQHE) is a transport anomaly that occurs whenever a two-dimensional electron system (2DES) in a strong perpendicular magnetic field has a gap for charged excitations at a fractional value of the Landau level (LL) filling factor.  Gaps at fractional filling factors can only be produced by electron-electron interactions. 
The FQHE has therefore been a rich playground for the study of strongly correlated phases of the electron liquid,
hosting a variety of exotic phenomena including fractional and non-Abelian quasiparticle statistics~\cite{Nayak} and 
electron-hole pair superfluidity~\cite{Eisenstein}.

Since its discovery~\cite{Tsui} more than three decades ago, 
the FQHE has been studied almost exclusively in the two-dimensional electron systems (2DESs) formed 
near GaAs/AlGaAs heterojunctions. 
Because of their small Zeeman to cyclotron energy ratio~\cite{Halperin}, 
the electron spin degree-of-freedom in the $N=0$ LL of the GaAs conduction band is often 
experimentally relevant, endowing the FQHE with ground and quasiparticle states that 
would not occur in the spinless fermion case~\cite{Sondhi}.   

The $N=0$ LL of monolayer graphene is nearly four-fold degenerate because of the presence of spin and valley degrees of freedom,
and is partially occupied over the filling factor range from $\nu= -2$ to $\nu=2$, opening the door to SU(4) 
manifestations of the FQHE. However, because graphene sheets on substrates generally have stronger disorder than 
modulation-doped GaAs/AlGaAs 2DESs, it has until recently not been possible to observe their fractional quantum Hall effects. Recent studies of high-quality graphene samples have started to clear 
the fog~\cite{Du,Bolotin,Ghahari,Dean,Feldman,Feldman2} however, and the view that has emerged is surprising.   
Experiments indicate that the 
graphene FQHE is stronger for $0 < |\nu| < 1$ than for $1 < |\nu| <2$, 
and that phase transitions between distinct states at the same $\nu$ occur as a function of magnetic field strength~\cite{Feldman,Feldman2}.
In this Letter we shed light~\cite{Abanin} on these trends by
using a variational approach to account for weak SU(4) symmetry breaking, and 
by constructing rules that allow SU(4) FQHE states in the 
range $0 < |\nu| <1$ to be generated starting from well known {\it seed} states in the range $1 < |\nu| < 2$.
Surprisingly, we find that in the absence of symmetry breaking terms the ground state at $|\nu|=1/3$ is {\it not} of the 
simple Laughlin type.

\noindent
{\em Hard-core SU(4) states}---We begin by considering the SU(4) invariant Coulomb-interaction model
in the $N=0$ LL.  It is convenient to define a filling factor measured from the empty $N=0$ LL:
$\tilde{\nu} \in[0,4]= 2+\nu$.  Progress can be achieved by starting
from $\tilde{\nu}\leq1$ zero-energy eigenstates of the $V_0$ hard-core model, in which only the
$m=0$ Haldane pseudo-potential is non-zero~\cite{Haldane}.
We will refer to these states as  {\em seed states} in the remainder of paper.
Note that the manifold of seed states is large and includes many states that are
not relevant at low energies.  However, our assumption is that the Coulomb interaction will select a ground state from among those states
on the basis of $m>0$ Haldane pseudo-potentials.
What is crucial for what follows is that seed states are not influenced by the short range interactions which break the SU(4) symmetry because they have zero probability for the spatial coincidence of particles.
They can be written as a product of the Vandermonde determinant and a SU(4) bosonic wavefunction, which forces them to have filling factors $\tilde{\nu}\leq1$~\cite{MacDonald,dassarmayang}.
Because of the Pauli exclusion principle, seed states include all the single component incompressible states like the Laughlin states~\cite{Laughlin}, single component composite Fermion states~\cite{Jain,Jainbook}, and Moore-Read states~\cite{Moore}. Several multicomponent states, like the spin-singlet Halperin state at $\tilde{\nu}=2/5$~\cite{Halperin}, also belong to this class.

We now demonstrate that many important incompressible states with $\tilde{\nu}\in(1,4]$ are simply related to 
$\tilde{\nu}\leq1$ seed states.  We first note that global particle-hole symmetry of the $N=0$ LL maps eigenstates with 
$\tilde{\nu}\in[0,2]$ to eigenstates at $4-\tilde{\nu}\in[2,4]$. This reduces our task to 
constructing states in $\tilde{\nu}\in(1,2]$. 
In the following we denote multicomponent states by a vector specifying the partial fillings of each non-empty component:
$(\nu_1,\cdots,\nu_k)$, with $\tilde{\nu}=\sum_i{\nu_i}$. (We require $\nu_i\geq\nu_{i+1}$ to avoid double counting
states that are related by a global SU(4) transformation.) 
Two simple mappings generate states in $\tilde{\nu}\in[1,2]$ from seed states in 
$\tilde{\nu}\in[0,1]$. The first is particle-hole conjugation restricted to two-components which maps $(\nu_1,\nu_2)$ 
to $(1-\nu_2,1-\nu_1)$~\cite{Abanin}. 
The second takes any seed wavefunction with three components or less, {\it i. e.} $(\nu_1,\cdots,\nu_k)$ with $k\leq3$, 
and multiplies it by the Vandermonde determinant of one of the empty components, 
producing a state with flavor composition $(1,\nu_1,\cdots,\nu_k)$. 
Particle-hole conjugation involving 3 components does not yield states 
that cannot be obtained by combining these two rules.

We will focus on the 
states at $\tilde{\nu} = p/3$, with $p=\{1,2,4,5\}$. The $\tilde{\nu}\leq1$ seed states are well known.
The ground state for $\tilde{\nu}=1/3$ is the Laughlin state which is an SU(4) ferromagnet.  At 
$\tilde{\nu}=2/3$, the single component particle-hole conjugate of the Laughlin state
competes with the two-component singlet state with flavor composition $(1/3,1/3)$ which
has lower Coulomb energy~\cite{Xie} and 
can be thought of as a composite fermion state with negative effective field~\cite{Jainbook}. 
At $\tilde{\nu}=4/3$, we obtain two competing states  with
flavor compositions $(1,1/3)$ and $(2/3,2/3)$, obtained by the two-component
particle-hole conjugation from $\tilde{\nu}=2/3$. 
These two states are well known from work on the FQHE of spinful fermions.  
However, at $\tilde{\nu}=5/3$ we obtain two states 
by acting on the seed states at $\tilde{\nu}=2/3$ with the second mapping.  These 
states have flavor composition $(1,2/3)$ and $(1,1/3,1/3)$.
The appearance of a three-component state at $\tilde{\nu}=5/3$
demonstrates that there is no reason to anticipate a simple relationship between  
$\tilde{\nu}$ and $2-\tilde{\nu}$ states in graphene.
The $(1,1/3,1/3)$ state has not previously 
been discussed as a possible $|\nu|=1/3$ ground state.

The energy of any state constructed via these mapping rules 
can be calculated provided the energy of the seed state is known. 
If the Coulomb energy per flux quantum of the seed state is $E_{\tilde{\nu}}$, then, the energy of the states obtained are respectively,
\be\label{Enu}
\begin{split}
E_{2-\tilde{\nu}}&=E_{\tilde{\nu}}+(1-\tilde{\nu}) 2 E_1,\\
E_{1+\tilde{\nu}}&=E_{\tilde{\nu}}+E_1,
\end{split}
\ee
\noindent where $E_1=-\sqrt{\pi/2} \ e^2/2 \epsilon l$ and $l$ is the magnetic length. 
This allows us to predict the energetic ordering of the $\tilde{\nu} = p/3$ states.
At $\tilde{\nu}=4/3$, the two-component particle-hole
conjugate of the singlet $(2/3,2/3)$ has lower Coulomb energy than the $(1,1/3)$ state. 
At $\tilde{\nu}=5/3$, Eqs.~\eqref{Enu} predict that $(1,1/3,1/3)$ 
has lower Coulomb energy than $(1,2/3)$. 
This observation is important, because  the state that has been thought to be experimentally realized is 
$(1,2/3)$~\cite{Feldman,Feldman2,Abanin}, not $(1,1/3,1/3)$.
We note that, although our discussion has been centered around the incompressible ground 
states, the mappings and Eqs.~\eqref{Enu} apply equally well to charged and neutral excited states 
generated from zero-energy eigenstates of the $V_0$ hard-core model.

\noindent
{\em Broken SU(4) symmetry}---
It has become clear from experimental~\cite{Young,Young2} and theoretical~\cite{Alicea,Herbut,jeil,kharitonov,Abanin} studies that short-range valley-dependent corrections to the long-range SU(4) symmetric 
Coulomb interactions play a significant role in determining the ground state of the quantum Hall ferromagnet
state realized at neutrality ($\tilde{\nu}=2$) in graphene. 
In this section we describe their influence on the $N=0$ fractional quantum Hall regime. 
The symmetry breaking interactions can be modeled as {\it zero-range} valley-dependent pseudo-potentials~\cite{kharitonov},
\begin{equation} 
\label{eq:ham}
H_a = \sum_{i<j,\sigma} \, V_{\sigma} \, \tau^{i}_{\sigma} \, |0\rangle_{ij}  {}_{ij}\langle0| \, \tau^{j}_{\sigma} 
\end{equation} 
\noindent where $\tau^{i}_{\sigma}$ is a Pauli matrix which acts on the valley degree of freedom 
of particle $i$, $\sigma=\{x,y,z\}$, $|0\rangle_{ij}  {}_{ij}\langle 0|$ projects the pair state of particles $i$ and $j$ onto relative angular momentum $0$, and $V_{\sigma}$ is a valley-dependent Haldane pseudopotential.   
Because conservation of total crystal momentum
implies that the number of electrons in each valley is conserved, we have 
$V_{x}=V_{y}\equiv V_{\perp}$. The system's weakly-broken SU(4) symmetry is 
therefore characterized by three parameters $V_{z}$, $V_{\perp}$, and by the 
Zeeman field strength $h$. The values of $V_{z}$ and $V_{\perp}$ are dependent on the component of 
magnetic field perpendicular to the graphene plane $B_\perp$, whereas the Zeeman strength
is determined by the total magnetic field, therefore, their relative strengths can be controlled by tilting the magnetic field away from the 2DES normal. 

We assume that the symmetry breaking terms are not strong enough to alter the Coulomb correlations of the 
SU(4) model states.  Much as in the case of standard magnetic systems, 
the role of the anisotropy terms is to select the $4$-component spinors assigned to wave function components.
Since more than one incompressible state might enjoy good Coulomb correlations 
at a given $\tilde{\nu}$, symmetry breaking terms will also alter the energy balance 
between these states. In order to compute the contribution to
total energy arising from the symmetry breaking terms,
we separate the spinors into those that are {\it completely} filled whose orbital wavefunction is a Slater determinant, 
and those that are {\it fractionally} filled whose orbital wavefunction is a hard-core model zero-energy eigenstate~\footnote{For some states, like $(2/3,2/3)$, a global particle-hole conjugation might be needed in order to conceptualize its fractionally filled spinors as having a hard-core wavefunction.}. The total anisotropy energy per flux quantum is

\begin{equation}\label{EI}
\epsilon_a=\frac{1}{2}{\rm tr}(P_i H^{HF}_i)+{\rm tr}(P_f H^{HF}_i)-\frac{h}{2} {\rm tr}(P_i \sigma_z),
\ee 

\noindent where $P_i=|\chi_1\rangle\langle\chi_1|+\cdots+|\chi_k\rangle\langle\chi_k|$ is the projector onto the 
completely filled spinors, 
$P_f=\nu_{k+1}|\chi_{k+1}\rangle\langle\chi_{k+1}|+\cdots+\nu_{4}|\chi_{4}\rangle\langle\chi_{4}|$ 
is a weighed projector onto fractionally filled spinors , $\sigma_z$ is a Pauli matrix acting on spin, and $h=g\mu_B B/2$. 
In Eq.~\eqref{EI} $H_i^{HF}$ is the anisotropy contribution to the 
Hartree-Fock quasi-particle Hamiltonian that one would obtain if there were no 
fractionally occupied components:
\be\label{HF}
H^{HF}_i=\sum_\sigma V_{\sigma} \left[{\rm tr}(P_i\tau_\sigma) \tau_\sigma-\tau_\sigma P_i\tau_\sigma\right]-h \sigma_z.
\ee
The spinors which appear in the 
projection operators are fixed by minimizing the anisotropy energy.
Equation~\eqref{EI} follows from the hard-core assumption, and from the following property of completely filled spinors:
\be
\hat{\rho}_m(r)|\Psi\rangle=\frac{1}{2 \pi l^2}|\Psi\rangle,
\ee  
\noindent where $\hat{\rho}_m(r)\equiv \hat{P}_{LLL}(\sum_i\delta(\hat{r}_i-r)|\chi_m\rangle_i{}_i\langle\chi_m|)\hat{P}_{LLL}$ is the particle density projected to the $m$-th completely filled spinor. 
These equations can be viewed as a generalization of the Hartree-Fock 
theory of integer quantum Hall ferromagnets. 
In particular, Eq.~\eqref{EI} reproduces the anisotropy energy expressions in Ref.~\cite{kharitonov} for the special case of neutral graphene, {\it i.e.} for $\nu_1=\nu_2=1$ and $\nu_3=\nu_4=0$. 
It also reproduces the expressions of Ref.~\onlinecite{Abanin} for the special case where the 
fractionally filled spinors are assumed to have canted antiferromagnetic order. 

Eq.~\eqref{EI} can also be used to compute anisotropy energy contributions to the
charge gaps. 
Assuming that quasiparticle states in the broken symmetry case evolve {\it adiabatically} from
SU(4) states, we label them by SU(4) quantum numbers.
Quasielectron-quasihole pair states can be labeled 
by integers which specify changes in the occupation numbers for 
each flavor relative to the incompressible ground state. 
Assuming that flavor flips involve only the fractionally filled and empty spinors, 
the integers satisfy $\delta N_{k+1}+\cdots+\delta N_4=0$~\footnote{Quasiparticles might involve flavor flips from the completely filled spinors. In the cases of interest in this letter, we bring back the problem into the current formulation by performing a global particle-hole transformation.}. 
We find that the gap of an incompressible state is the SU(4) Coulomb gap plus the following correction:

\be\label{Deltaa}
\Delta_a=\sum_{j=k+1}^{4}\delta N_j \langle\chi_{j}|H^{HF}_i|\chi_{j}\rangle.
\ee

\noindent {\em Ground states and gaps at $\tilde{\nu}=p/3$}---
The hard-core seed states at $\tilde{\nu}=1/3$ and $\tilde{\nu}=2/3$ do not experience 
the short range valley-dependent interactions~\cite{Feldman2,Abanin}. 
At $\tilde{\nu}=1/3$ we therefore expect a fully spin polarized Laughlin state, with a remnant valley SU(2) symmetry. 
The quasiparticles are therefore expected to be large valley skyrmions~\cite{Sondhi,Moon}. The gap is expected to be reduced by a factor of approximately $5$, relative to the 
single-component case, to $\Delta_{1/3}^{sky} \approx 0.023 \ e^2/\epsilon l$
\cite{Sondhi,Moon,Palacios}, possibly explaining why it is unobservable in suspended graphene samples~\cite{Dean,Feldman,Feldman2,Abanin,Hunt}. 
At $\tilde{\nu}=2/3$, we expect a fully spin polarized valley-singlet state. 
Two types of quasiparticles might be relevant at this filling fraction. 
In the absence of Zeeman terms, quasiparticles could lower their Coulomb energy by 
making flavor flips into the completely empty spinors. This is the behavior found for composite fermion wavefunctions at
$\tilde{\nu}=2/5$~\cite{Toke2}. A numerical study of SU(4) flavor reversed quasiparticles would be needed to quantitatively assess this scenario at $\tilde{\nu}=2/3$. At higher fields one 
would recover the picture of fully spin polarized quasiparticles in the SU(2) valley space.
The gap would then be~\cite{Xie} $\Delta_{2/3}=0.0784 e^2/\epsilon l$.~\footnote{These quasiparticles do not involve valley flips, since additional flips in the background of a singlet state tend to increase the Coulomb energy of the quasiparticles, contrary to the situation for polarized states~\cite{Vyborny}.}.  

%%%%% FIG 1 %%%%%
\begin{figure}[t]
\includegraphics[width=3.3in]{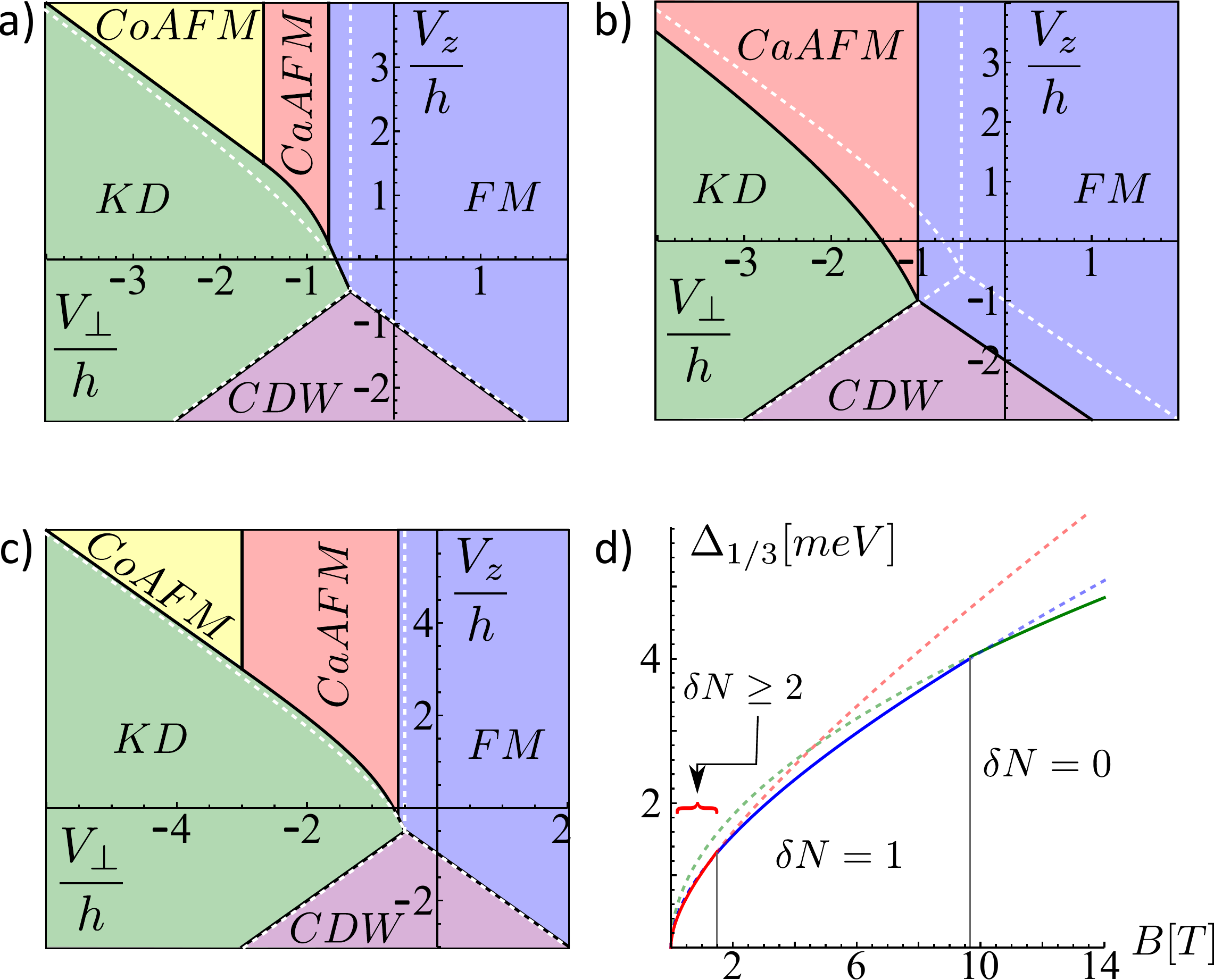}
\caption{(color online). Phase diagrams for: a) $(1,1,2/3)$, b) $(1,1,1/3,1/3)$, c) $(1,1,1/3)$. 
The dashed lines are the boundaries of the integer quantum Hall ferromagnet states
at neutrality~\cite{kharitonov}. 
The valley-dependent interaction parameters are believed to place graphene in the collinear antiferromagnet (CoAFM) 
region for panels a) and c), and in the canted antiferromagnet (CaAFM) region for b). 
FM, KD and CDW correspond to ferromagnet, Kekule-distortion and charge-density-wave phases respectively~\cite{supp}.
d) Field dependence of the gap for the Laughlin-type state $(1,1,1/3)$, $\delta N$ indicates the number of spin-flips of the corresponding quasielectron-quasihole pair.}
\label{fig1}
\end{figure}
%%%%%%%%%%%%%%%%%

Anisotropy has a greater impact for $\tilde{\nu}=\{4/3,5/3\}$.  
At $\tilde{\nu}=4/3$ we have two candidate incompressible states, namely $(1,1/3)$ and $(2/3,2/3)$. To discuss their competition, it is convenient to perform a global particle-hole transformation to the states 
$(1,1,2/3)$ and $(1,1,1/3,1/3)$ respectively. 
An analysis of the possible ordered phases leads to the phase diagram in Fig.~\ref{fig1}~\cite{supp}. 
Experiments suggest canted antiferromagnetic order at $\tilde{\nu}=2$~\cite{Young2}, and
are~\cite{Abanin} consistent with $V_\perp/h\sim-10$. 
According to the phase diagrams in Fig.~\ref{fig1}, this 
would imply that the $(1,1,2/3)$ state is a collinear antiferromagnet (CoAFM) in perpendicular field measurements, 
whereas $(1,1,1/3,1/3)$ is a canted antiferromagnet (CaAFM). 
We estimate that the critical field for the transition between $(1,1,2/3)$ and $(1,1,1/3,1/3)$ states is  
\be\label{Bc2/3}
B_c=\frac{1}{(1-h/|V_\perp|)^2}\left(\frac{\delta \epsilon^c_{2/3} }{h}\right)^2,
\ee 
\noindent where $\delta \epsilon^c_{2/3}$ is the Coulomb energy difference per-particle between the single 
component state and the singlet at $\tilde{\nu}=2/3$ and all the quantities on the right hand side of this
equation are understood to be evaluated at 1T.  
Exact diagonalization studies find that $\delta \epsilon^c_{2/3}\approx0.009e^2/\epsilon l$~\cite{Xie,Davenport,Niemel,Vyborny}. 
(Composite Fermion trial wavefunctions significantly underestimate this difference, 
although they correctly predict the ground state to be a singlet~\cite{Wu}.) 
In a SU(2) system like GaAs with symmetry broken only by Zeeman, 
the transition at $\tilde{\nu}=2/3$ occurs at $B_c=(\delta \epsilon^c_{2/3}/h)^2$. 
Eq.~\eqref{Bc2/3} reduces to this expression for $h\ll|V_\perp|$
because the anisotropy energy difference between the 
CoAFM and CaAFM states is dominated by Zeeman energies in this limit~\cite{supp}. 
We therefore obtain that $B_c=4.7$T for $|V_\perp|/h=10$ and 
$B_c=6$T for $|V_\perp|/h=5$~\footnote{In graphene $h/(e^2/\epsilon l) \sim 0.001  \epsilon \sqrt{B[{\rm T}]}$. We qualitatively include the impact of screening by using the RPA value of the dielectric constant of suspended graphene $\epsilon = 1+\pi \alpha/2 \approx 4.6$~\cite{Fogler}.},
in agreement with experiment~\cite{Feldman2}. An analysis of the gaps for the states at $\tilde{\nu}=4/3$ indicates that 
the quasiparticles involve a few flavor flips~\cite{longpaper}, in analogy with GaAs~\cite{Vyborny}. 

We will now discuss the competition at $\tilde{\nu}=5/3$ between the three-component state $(1,1/3,1/3)$ and the two-component state $(1,2/3)$. To the best of our knowledge previous theoretical studies 
have assumed that the incompressible state at $\tilde{\nu}=5/3$ is the $(1,2/3)$ Laughlin-type state, 
although, the exact diagonalization study of Ref.~\cite{Papic}, in which 
a finite Zeeman field was needed to stabilize the Laughlin state at $\tilde{\nu}=7/3$, did provide a contrary hint. 
Note that $\tilde{\nu}=7/3$ is the global particle-hole conjugate of $\tilde{\nu}=5/3$. 
As previously shown, the singlet-type state $(1,1/3,1/3)$ has lower Coulomb energy. 
When anisotropy is included we find that the fully filled spinor of  $(1,1/3,1/3)$ is $|K,\uparrow\rangle$,  
while the fractionally-filled are $|K',\downarrow\rangle$, and $|K',\uparrow\rangle$~\cite{supp}. The anisotropy energy per flux quantum of this state
exceeds that of the $(1,2/3)$ CoAFM state by $2(|V_\perp|-h)/3$. 
We therefore predict a transition from $(1,1/3,1/3)$ to $(1,2/3)$ at the critical field, 

\be
B_c=\left(\frac{\delta \epsilon^c_{2/3} }{|V_\perp|-h}\right)^2,
\ee

\noindent where all quantities in the right hand side are evaluated at 1T. 
For $|V_{\perp}|/h = 10$ we obtain $B_c=0.045$T.  
(For $|V_{\perp}|/h = 5$ we obtain $B_c=0.23$T.) 
The transition field is small because $|V_{\perp}|\gg h$.
This is consistent with the absence of an experimental
transition in the field range where the FQHE is clearly observable~\cite{Feldman2}. 
Our estimates indicate, however, that this critical field increases with tilted magnetic field,
making the realization of the three component $\nu=1/3$ state an experimental 
possibility.~\cite{longpaper}. 

Finally, we apply our formalism to determine the charge gaps of the particle-hole equivalent
Laughlin-like states at $\tilde{\nu}=5/3$ and $\tilde{\nu}=7/3$, namely $(1,2/3)$ and $(1,1,1/3)$. 
In the perpendicular field configuration these states are expected to be in the CoAFM phase~\cite{supp}. 
For this state there are two kinds of quasiparticles involving flavor flips. 
The first involves flips from the completely filled spinors. 
These quasiparticles have lower Coulomb energy, but considerably larger anisotropy energy and are thus likely irrelevant in
experiment~\cite{supp}. We will focus on the second kind, which involve flips between the fractionally filled and the empty spinors.
For the CoAFM state, $(1,1,1/3)$, we can choose the completely filled spinors to be 
$|K,\uparrow\rangle$, $|K',\downarrow\rangle$, and the $1/3$ filled spinor to be $|K',\uparrow\rangle$. 
The quasiparticles can lower their energy by flavor flips from the spinor, $|K',\uparrow\rangle$, into the unoccupied 
spinor $|K,\downarrow\rangle$. The anisotropy contribution to the gap from Eq.~\eqref{Deltaa} per flavor flip is simply $2h$, the conventional single spin-flip Zeeman gap. 
This is analogous to the situation of GaAs at $\tilde{\nu}=1/3$, where one expects the quasiparticles of the Laughlin 
state to involve a few spin flips up to magnetic fields $\sim10$T~\cite{Chakraborty,Rezayi,Palacios,Wojs,Dethlefsen}. 
Hence, it is likely that the quasiparticles of the $\tilde{\nu}=\{5/3,7/3\}$ states in graphene involve a few spin flips as well. 

Let us assess this scenario quantitatively. 
The conventional Coulomb gap of the Laughlin state without flavor flips is $\Delta_{1/3}^0 \approx 0.1036 \ e^2/\epsilon l$~\cite{Fano}. 
The gap for a single flip corresponds to a spin-flipped quasielectron and a no-flip quasihole pair, 
and it is about $\Delta_{1/3}^{1} \approx 0.075 \ e^2/\epsilon l$~\cite{Chakraborty,Rezayi,Palacios,Dethlefsen}. 
The gap for two flavor flips, $\Delta_{1/3}^{2}$, is known with less accuracy, but can be estimated to be lower 
than $\Delta_{1/3}^{1}$ by about $0.01 \ e^2/\epsilon l$~\cite{Palacios,Wojs,Dethlefsen}, and it is expected to 
correspond to a single spin-flipped quasielectron and single spin-flipped quasihole pair. 
The predicted gap behavior is depicted in Fig.~\ref{fig1}, and
is in good agreement with experiment~\cite{Feldman, Feldman2}. 
Figure~\ref{fig1} indicates that for most of the range probed in 
Refs.~\cite{Feldman, Feldman2} the relevant quasiparticles involve a single spin flip.

In summary, we have developed a method to construct multicomponent incompressible and 
quasiparticles states for the $N=0$ LL of graphene, starting from $V_0$ hard-core model seed states. 
We have provided simple variational formulas to determine how the short-range valley dependent 
interactions select the broken symmetry ground states and influence
their gaps. We have applied this formalism to study the ground states and quasi-particles
at $\tilde{\nu}=p/3$, revealing a previously unnoticed state with lower Coulomb energy than the Laughlin state at $\tilde{\nu}=5/3$.   

We thank S. Davenport, R. Hegde, K. V\'yborn\'y, C. T\H oke, and A. W\'ojs for valuable interactions. 
This work was supported by the DOE Division of Materials Sciences and Engineering under grant DE-FG03-02ER45958,
and by the Welch foundation under grant TBF1473.
IS acknowledges support from a Graduate School Named Continuing Fellowship at the 
University of Texas at Austin.

\newpage

\section{Supplementary Material} 

\section{ordered phases for two component states in the filling factor range $|\nu|<1$}

In this section we consider the subset of incompressible states in the filling factor 
range $|\nu|<1$ which can be viewed as arising directly from quasiparticles formed formed in the $\nu=0$ 
quantum Hall ferromagnet ground state.  For $\tilde{\nu}\in(1,2)$ these states partially occupy only two spinors. 
The equivalent particle-hole conjugate states in 
$\tilde{\nu}\in(2,3)$ fully occupy two spinors and partially occupy two other spinors. 
Without loss of generality we will describe only the $\tilde{\nu}\in(2,3)$ case for which the flavor composition is $(1,1,\nu_3,\nu_4)$. 
Let us call the fully occupied spinors $|\chi_1\rangle$ and $|\chi_2\rangle$ and the partially occupied spinors $|\chi_3\rangle$ and $|\chi_4\rangle$. We assume that the spinors that minimize the energy do not have valley-spin entanglement, {\it i.e.} they can be written as $|\chi_i\rangle=|{\mathbf t_i}\rangle\otimes|{\mathbf s_i}\rangle\equiv|{\mathbf t_i},{\mathbf s_i}\rangle$, where ${\bf t}_i$ denotes a unit vector in the valley Bloch sphere, and ${\bf s}_i$ denotes a unit vector in the spin Bloch sphere, in analogy with the quantum Hall ferromagnet at neutrality~\cite{kharitonov}.  Given this assumption, one finds that the states which minimize the anisotropy energy (Eq.(3) in the main text), can be seperated into two classes: spin-ordered phases and valley-ordered phases. The spin ordered phases have spinors,

\be
\begin{split}
&|\chi_1\rangle=|K,{\mathbf s}_K\rangle, \; \; \; \; |\chi_2\rangle=|K',{\mathbf s}_{K'}\rangle,\\
&|\chi_3\rangle=|K,-{\mathbf s}_K\rangle, \;  |\chi_4\rangle=|K',-{\mathbf s}_{K'}\rangle.
\end{split}
\ee 

\noindent Their anisotropy energy per flux quantum can be shown to be,

\be
\begin{split}
\epsilon_a=&-V_\perp (1-\nu) \mathbf{s}_K\cdot \mathbf{s}_{K'}-V_z-V_\perp (1+\nu)\\
&-h(1-\nu_3)s^z_K-h(1-\nu_4)s^z_{K'},
\end{split}
\ee

\noindent where $\nu=\nu_3+\nu_4$. This equation is equivalent to Eqs.(18)-(20) of Ref.~\onlinecite{Abanin} up to an overall constant that arises from particle-hole conjugation. Within the spin ordered phases the state that minimizes the energy depends solely on the ratio $V_\perp/h$. Three different spin ordered phases are found depending on the value of this ratio. Without loss of generality we assume in the remainder that $\nu_3\geq\nu_4$ and $h \geq 0$. First we have a collinear antiferromagnet (CoAFM), where the spin orientations are collinear with the Zeeman field axis, ${\bf s}_K=-{\bf e}_z$ and ${\bf s}_{K'}={\bf e}_z$. This phase is stable for $V_\perp<0$ and,

\be\label{gamma1}
\frac{|V_\perp|}{h}\geq\frac{(1-\nu_3)(1-\nu_4)}{(1-\nu)(\nu_3-\nu_4)},
\ee
 
\noindent and has energy $\epsilon_a=-V_{z}-2\nu V_\perp-h(\nu_3-\nu_4)$. Second we have a canted antiferromagnet (CaAFM), where the spin orientations are canted away from the Zeeman field axis in opposite directions and with different canting angles in each valley in general. This phase is stable for $V_\perp<0$ and,

\be\label{gamma2}
\frac{(1-\nu_3)(1-\nu_4)}{(1-\nu)(\nu_3-\nu_4)}\geq\frac{|V_\perp|}{h}\geq\frac{(1-\nu_3)(1-\nu_4)}{(2-\nu)(1-\nu)},
\ee

\noindent the energy and canting angles of the spinors in each valley are,

\be\label{thetaK}
\begin{split}
\epsilon_a=&|V_\perp|(1+\nu)-\frac{|V_\perp|}{2}(1-\nu) \left(\frac{1-\nu_4}{1-\nu_3}+\frac{1-\nu_3}{1-\nu_4}\right)\\
&-\frac{h^2}{2|V_\perp|}\frac{(1-\nu_3)(1-\nu_4)}{(1-\nu)}-V_{z},\\
s^z_K=&\frac{h(1-\nu_4)}{2 |V_\perp| (1-\nu)}+\frac{|V_\perp| (1-\nu)}{2 h (1-\nu_4)}\Biggl[1-\biggl(\frac{1-\nu_4}{1-\nu_3}\biggr)^2\Biggr],
\end{split}
\ee

\noindent and $s^z_{K'}$  can be obtained from above expression by switching labels $3\leftrightarrow 4$. The third and last spin-ordered phase is the ferromagnet (FM), where both spins point along the Zeemann axis ${\bf s}_K={\bf s}_{K'}={\bf e}_z$, and is stable in the remaining range of $V_\perp/h$, and has energy $\epsilon_a=-V_{z}-2V_\perp-h(2-\nu)$. An important special case is when the two valleys are equally filled, {\it i.e.} $\nu_3=\nu_4$.   As this limit is approached 
the boundary to the CoAFM goes to infinity and only the CaAFM and FM phases are present. 
For $\nu_3=\nu_4$ the valleys have canting angles of equal magnitude and opposite sign with respect to the Zeeman axis. 
Note that all transitions between the spin-ordered phases are continuous.

The second class of states, the valley ordered states, occupy spinors 

\be
\begin{split}
&|\chi_1\rangle=|{\bf t},\uparrow\rangle, \;\;\;\;\;\;  |\chi_2\rangle=|{\bf t},\downarrow\rangle,\\
&|\chi_3\rangle=|-{\bf t},\uparrow\rangle, \; |\chi_4\rangle=|-{\bf t},\downarrow\rangle.
\end{split}
\ee 

\noindent One finds two valley-ordered phases that minize the anisotropy energy. First the charge-density-wave (CDW) phase with ${\bf t}=\pm{\bf e}_z$, where the north/south poles of the Bloch sphere designate valleys $K/K'$. The CDW phase has energy $\epsilon_a=V_{\perp}(1-3\nu)-\nu V_z-h(\nu_3-\nu_4)$. Second the Kekule-distortion (KD) phase with $t_z=0$, and energy $\epsilon_a=V_{z}(1-2\nu)-2\nu V_\perp-h(\nu_3-\nu_4)$. The KD-CDW phase transition is first order
 and occurs along the line $V_\perp=V_z$. It terminates at the multicritical point $V_\perp=V_z=-h(1-\nu_3)/(2(1-\nu))$, where three phases coexist (KD-CDW-FM) for $\nu_3>\nu_4$ and four phases coexist (KD-CDW-FM-AFM) for $\nu_3=\nu_4$.
Phase diagrams illustrating the dependence of pseudospin phase on symmetry breaking interaction parameters 
of several incompressible states are provided 
in Fig.[1] of the main text.  These diagrams are constructed by comparing the 
anisotropy energies of the phases discussed above.  

\section{phases of the three component state at $\tilde{\nu}=5/3$}

%%%%% FIG 1 %%%%%
\begin{figure}
\includegraphics[width=2.7in]{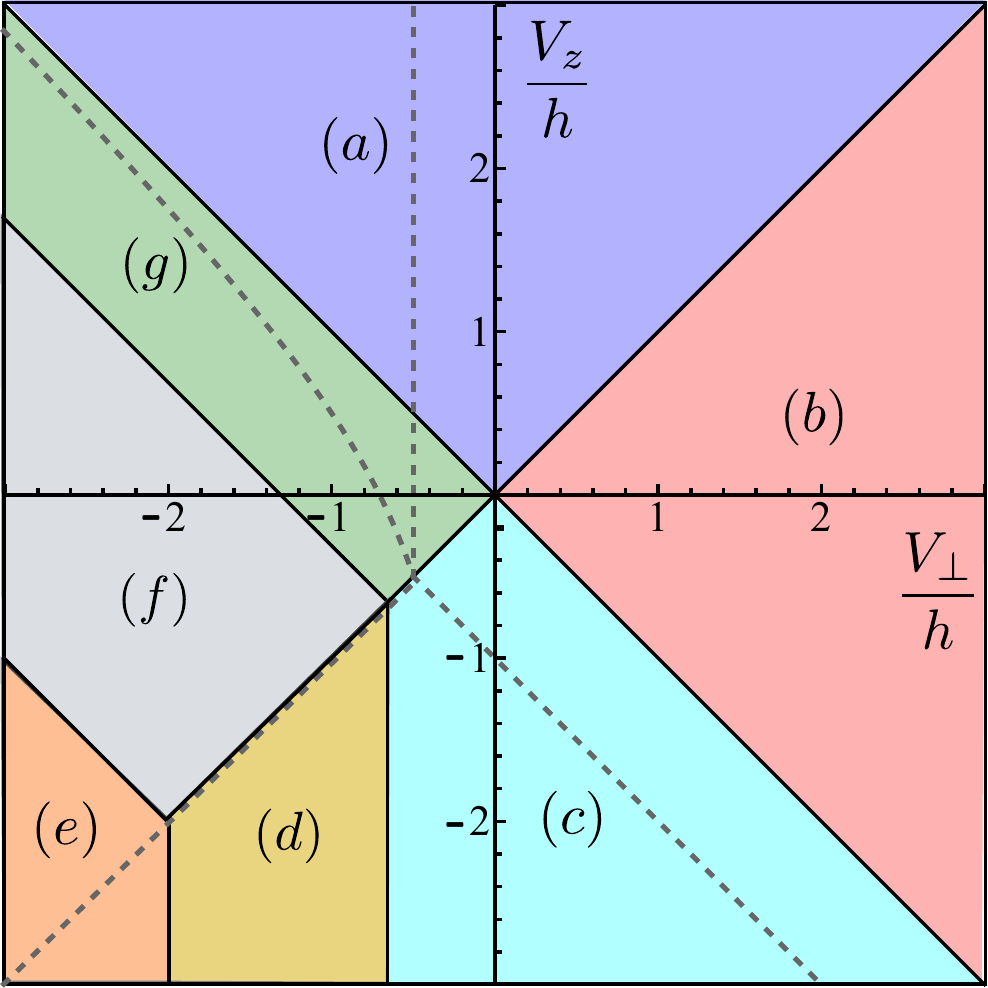}
\caption{(Color online) Phase diagram of three component state $(1,1/3,1/3)$. See text for description of the labels. The dashed lines correspond to the phase boundaries for the integer quantum Hall ferromagnet realized at neutrality~\cite{kharitonov}.}
\label{fig1supp}
\end{figure}
%%%%%%%%%%%%%%%%%

At $\tilde{\nu}=5/3$, in addition to the two-component Laughlin-type state which falls within the classes described in 
previous section, there is a three component incompressible state with flavor composition $(1,1/3,1/3)$. 
We will discuss in this section the possible symmetry breaking patterns that minimize the anisotropy energy of this state. We assume again that these states have no valley-spin entanglement, and thus that
the occupied spinors have the form $|\chi_i\rangle=|{\mathbf t_i}\rangle\otimes|{\mathbf s_i}\rangle\equiv|{\mathbf t_i},{\mathbf s_i}\rangle$, where ${\bf t}_i$ denotes a unit vector in the valley Bloch sphere, and ${\bf s}_i$ denotes a unit vector in the spin Bloch sphere.  With this restriction, we find {\it seven} phases that minimize the anisotropy energy for different values of the symmetry breaking parameters. They fill spinors: (a) $|K,\uparrow\rangle$, $|K',\downarrow\rangle$, $|K',\uparrow\rangle$; (b) $|{\mathbf t}_\perp,\uparrow\rangle$, $|-{\mathbf t}_\perp$, $\uparrow\rangle$, $|-{\mathbf t}_\perp,\downarrow\rangle$; (c) $|K,\uparrow\rangle,|K,\downarrow\rangle,|K',\uparrow\rangle$; (d) $|K,{\mathbf s}_1\rangle,|K',{\mathbf s}_2\rangle,|K,-{\mathbf s}_1\rangle$; (e) $|{\mathbf t},\uparrow\rangle,|K,\downarrow\rangle,|K',\downarrow\rangle$; (f) $|{\bf t}_\perp,{\mathbf s}_1\rangle,|-{\bf t}_\perp,{\mathbf s}_2\rangle,|{\bf t}_\perp,-{\mathbf s}_1\rangle$; (g) $|{\mathbf t}_\perp,\uparrow\rangle,|-{\mathbf t}_\perp,\uparrow\rangle,|{\mathbf t}_\perp,\downarrow\rangle$. In this listing the first spinor is understood to be fully filled and the other two to be fractionally filled, $\mathbf{t}_\perp$ is a unit vector on the equator of the valley Bloch sphere, $\mathbf{t}$ is an arbitrary unit vector on the valley Bloch sphere, and $\{\mathbf{s}_1, \mathbf{s}_2\}$ are unit vectors in the spin Bloch sphere. Any of the listed states with definite valley numbers is understood to have the $\mathbb{Z}_2$ valley interchange symmetry $K\leftrightarrow K'$,
and we have listed only one of its realizations. The anisotropy energy per flux quantum of these phases are,

\be
\begin{split}
(a)& \ \epsilon_a=-\frac{2}{3} (V_z+V_\perp)-h,\\
(b)& \ \epsilon_a=-\frac{1}{3} (V_z+3 V_\perp)-h,\\
(c)& \ \epsilon_a=-\frac{2}{3} V_\perp-h,\\
(d)& \ \epsilon_a=\frac{1}{12}V_\perp+\frac{1}{3}\frac{h^2}{V_\perp},\\
(e)& \ \epsilon_a=-\frac{h}{3},\\
(f)& \ \epsilon_a=\frac{1}{24}(V_z+V_\perp)+\frac{2}{3}\frac{h^2}{(V_z+V_\perp)},\\
(g)& \ \epsilon_a=-\frac{1}{3} (V_z+V_\perp)-h.\\
\end{split}
\ee 

Phases (d) and (f) have spins canted away from the Zeeman field. 
The projection of the spins along the Zeeman axis are

\be
s_z=\frac{h}{2 |V_\perp|}+\frac{3 |V_\perp|}{8 h}, \ \ s_{2z}=\frac{h}{|V_\perp|}-\frac{3 |V_\perp|}{4 h}.
\ee 

\noindent for phase (d) and

\be
\begin{split}
& s_z=\frac{h}{|V_z+V_\perp|}+\frac{3 |V_z+V_\perp|}{16 h},\\
& s_{2z}=\frac{2 h}{|V_z+V_\perp|}-\frac{3 |V_z+V_\perp|}{8 h}.
\end{split}
\ee

\noindent for phase (f).  

The phase diagram obtained by comparing the energies of these 
states depicted in Fig.~\ref{fig1supp}. 
The dashed lines in Fig.~\ref{fig1supp} depict the boundaries of the ordered phases of the integer quantum Hall ferromagnet described in Ref.~\cite{kharitonov}. An important observation is that, given that the ground state of the integer quantum Hall ferromagnet is likely to be in the canted antiferromagnetic phase~\cite{Young2} and that $V_\perp \sim - 10 h$~\cite{Abanin}, the three component state that competes with the two component Laughlin-type $(1,2/3)$ collinear antiferromagnet state is likely to be in phase (a). This is true except for an extremely small region in the $V_\perp,V_z,h$ parameter space close to the boundary between the (g) and (a) phase in Fig.~\ref{fig1supp} (see also Fig.1(c) of the main text).

\section{Integer flavor flip quasiparticles at the Laughlin type state at $\tilde{\nu}=7/3$}

As discussed in the main text the two-component Laughlin type state at $\tilde{\nu}=5/3$, {\it i.e.} $(1,2/3)$ is not the ground state in the absence of symmetry breaking terms, and instead the three component state $(1,1/3,1/3)$ has lower Coulomb energy. This statement also applies 
at $\tilde{\nu}=7/3$ because of the global particle-hole symmetry.  
We therefore expect the state $(1,2/3,2/3)$ to have lower Coulomb energy than $(1,1,1/3)$. However, sufficiently strong Zeeman or $V_\perp$ anisotropy terms will make $(1,1,1/3)$ have lower energy.
We believe that the state observed in experiments on suspended graphene samples in 
graphene is likely the $(1,1,1/3)$ state. 
For $V_\perp<0$ and $|V_\perp|>3h$ the phase that minimizes the anisotropy energy
of $(1,1,1/3)$ is the collinear antiferromagnet as explained 
in the first section of this supplement. 
This collinear antiferromagnetic Laughlin-like state is likely the one realized in the experiments of Refs.~\cite{Feldman,Feldman2}. 

We would like to determine what type of quasiparticles give rise to the charge gap of this state. These quasiparticles will
generally involve flavor flips. There are two types of flavor flips which is convenient to distinguish. 
The first kind is more easily conceptualized for the state $(1,1,1/3)$. For the collinear antiferromagnetic order the completely filled spinors can be chosen to be $|K\uparrow\rangle$ and $|K'\downarrow\rangle$. The neutral quasiparticle-quasihole pairs can involve flips from the $1/3$ filled spinor, i.e. $|K'\uparrow\rangle$, into the empty spinor $|K\downarrow\rangle$. The anisotropy energy contribution to this gap reduces simply to the Zeeman gap, $2h$, per flavor-flip. In particular, as discussed in the text, the flavor flipped quasi-electron and no flip quasi-hole are
expected to be the lowest energy excitations for most of the magnetic field range explored in Refs.~\cite{Feldman,Feldman2}. 
The gap associated with these excitations is expected to be $\Delta^1_{1/3}=0.075 e^2/\epsilon l+2h$, where the Coulomb energy has been extracted from exact diagonalization studies extrapolated to the 
thermodynamic limit~\cite{Rezayi,Palacios,Wojs,Dethlefsen}. 

A second kind of quasiparticle-quasihole pair associated with the state $(1,1,1/3)$, would involve flavor flips from either of the completely filled spinors into the $1/3$ filled spinor~\footnote{Quasiparticles involving a flip from one of the completely filled spinors into the completely empty spinor would have an associated Coulomb gap $\sqrt{\pi/2}e^2/\epsilon l$, which is considerably larger than those here considered and hence unlikely to be lowest energy charged excitations.}. These quasiparticles are more easily conceptualized in the particle-hole mirror state $(1,2/3)$, where they appear as involving flavor flips from the $2/3$ filled spinor into the completely empty ones. For $(1,2/3)$ if we choose the completely filled spinor to be $|K\uparrow\rangle$, the partially filled spinor would be $|K'\downarrow\rangle$ in the collinear antiferromagnetic phase. Since there are two completely empty spinors, there are two-types of flavor flips. In the first one we remove an electron from $|K'\downarrow\rangle$ and place it into  $|K'\uparrow\rangle$. Applying Eq.(6) from the main text one finds that these quasiparticles would have a gap $\Delta^{1}_{2/3}\approx0.051 e^2/\epsilon l-2(V_\perp+h)$. For the second type of flavor flip we remove an electron from $|K'\downarrow\rangle$ and place it into  $|K\downarrow\rangle$, these quasiparticles would have a gap $\Delta^{1}_{2/3}\approx0.051 e^2/\epsilon l+2 V_z$. We have obtained the Coulomb gap for a quasiparticle-quasihole pair involving a single flavor flip from the exact diagonalization studies of Ref.~\onlinecite{Vyborny}, which are not extrapolated to thermodynamic limit and thus might contain finite size effects. Reference~\onlinecite{Vyborny} found that the charged gap for a single spin-flip is associated with a single spin-flip quasielectron and a no spin-flip quasihole. Note that in the absence of symmetry breaking terms these quasi-particles are expected to have lower energy than those discussed in the previous paragraph. We believe this is a natural explanation for the finding in Ref.~\onlinecite{Papic}
that there is an intermediate regime in which the lowest energy excitations of the Laughlin like state $(1,1,1/3)$ 
involve flips from the completely filled spinors.

However, it is unlikely that the latter quasiparticles play a role at magnetic fields where the FQHE is observable. The reason is the relatively large anisotropy energy cost they involve. The critical fields at which the two types of charge gaps of the second kind 
of flavor flipped quasiparticle equals the first kind are,

\be
\begin{split}
& B_c=\left[\frac{0.024 e^2}{(2|V_\perp|-4h) \epsilon l}\right]^2,\\
& B_c=\left[\frac{0.024 e^2}{2 V_z \epsilon l }\right]^2,\\
\end{split}
\ee 

\noindent where the quantities in the right side are understood to be evaluated at $1$T. For the first critical field one obtains, $B_c=0.1$T for $V_\perp=-10 h$. One obtains $B_c=0.74$T for $V_\perp=-5 h$. The second critical fields is expected to be even smaller because the stability of collinear antiferromagnetic states requires $V_z\geq|V_\perp|$.

\end{document}